\documentclass[reprint,nofootinbib,superscriptaddress,amsmath,amssymb,aps,prl]{revtex4-1}
\usepackage{graphicx}
\usepackage{rotating}
\usepackage{dcolumn}
\usepackage{bm}
\usepackage{color}
\usepackage{mathptmx, textcomp}
\usepackage[latin1]{inputenc}
\usepackage{braket}
\usepackage[colorlinks,linkcolor=magenta,anchorcolor=cyan,citecolor=blue]{hyperref}
\usepackage[multiple]{footmisc}

\def\be{\begin{equation}}
\def\ee{\end{equation}}
\def\ba{\begin{eqnarray}}
\def\ea{\end{eqnarray}}

\newcommand{\eq}[1]{(\ref{#1})}

\def\q{\theta}         \def\d {\delta} \def\f {\phi}       \def\x {\xi}    \def\m {\mu} \def\pd {\partial}   
  \def\Y {\Psi}      \def\F {\Phi}      \def\grad{\nabla}\def\.{\cdot}
\def\math {\mathcal}
\begin{document}
\title{Comment on ``Constraints on Low-Energy Effective Theories from Weak Cosmic Censorship''}
\author{Jie Jiang}
\email{jiejiang@mail.bnu.edu.cn}
\affiliation{Department of Physics, Beijing Normal University, Beijing 100875, China\label{addr2}}
\author{Aofei Sang}
\email{202021140021@mail.bnu.edu.cn}
\affiliation{Department of Physics, Beijing Normal University, Beijing 100875, China\label{addr2}}
\author{Ming Zhang}
\email{Corresponding author. mingzhang@jxnu.edu.cn}
\affiliation{Department of Physics, Jiangxi Normal University, Nanchang 330022, China}
\date{\today}

\begin{abstract}
Recently, it was argued in [Phys. Rev. Lett. {\bf126}, 031102 (2021)] that the WCCC can serve as a constraint to high-order effective field theories. However, we find there exists a key error in their approximate black hole solution. After correcting it, their calculation cannot show the ability of WCCC to constrain the gravitational theories.
\end{abstract}
\maketitle

Weak cosmic censorship conjecture (WCCC) \cite{Penrose:1969pc} is a basic principle that guarantees the predictability of spacetime. One critical scientific question is whether the WCCC can give a new constraint to the gravitational theory. Recently, an attempt to this question was given in Ref. \cite{Chenprl}. Using Sorce-Wald's method \cite{SW} under the first-order approximation, the authors showed that the WCCC fails for some possible generations; thus they argued that the WCCC provides a constraint to the high-order low-energy effective theories(EFT). However, after examining their letter, we found that key errors occur in their approximate black hole solution. We will clarify this issue and show that their discussion cannot give the constraint to the high-order theories.

The Lagrangian of the EFT considered in \cite{Chenprl} is given by
\ba\begin{aligned}\label{L1}
\math{L}&=\frac{1}{2} R-\frac{1}{4}F_{ab}F^{ab}+c_1 R^2+c_2 R_{ab}R^{ab}+c_3 R_{abcd}R^{abcd}\\
&+c_4 R F_{ab}F^{ab}+c_5 R_{ab}F^{ac}F^b{}_c+c_6 R_{abcd}F^{ab}F^{cd}\\
&+c_7 F_{ab}F^{ab}F_{cd}F^{cd}+c_8 F_{ab}F^{bc}F_{cd}F^{da}
\end{aligned}\ea
where $c_i$'s are some coupling constants which are treated as small parameters in the calculations. The equation of motion (EOM) is given by
\ba\begin{aligned}\label{eom1}
H^{ab}=0\,,\quad\quad\grad_a E_F^{ab}&=0\,,
\end{aligned}\ea
in which $$H^{ab}=E_R^{cde(a}R_{cde}{}^{b)}+2\grad_c\grad_d E_R^{(a|c|b)d}-E_F^{c(a}F^{b)}{}_c-\frac{1}{2}g^{ab}\math{L}\,,$$
with $E_R^{abcd}=\pd\math{L}/\pd R_{abcd}$ and $E_F^{ab}=\pd \math{L}/\pd F_{ab}$.

First, we reexamine the solution given by Eqs. (6) and (7) in \cite{Chenprl}. With a straightforward calculation, it is easy to check
\ba\begin{aligned}
\grad_a E_F^{ab}(dt)_b&=\frac{2q^3}{r^7}[c_2+4c_3+10c_4+3(c_5+c_6)]+\math{O}(c_i^2)\quad \\
&\neq 0\,,
\end{aligned}\ea
which means that there are some errors in the solution given by Ref. \cite{Chenprl}. We start with the most general spherically symmetric static metric
\ba\begin{aligned}
ds^2=-f(r)dv^2+2\m(r) dv dr+r^2(d\q^2+\sin^2\q d\f^2)\,,
\end{aligned}\ea
and the Maxwell field
\ba\begin{aligned}
\bm{A}=\Y(r) dv\,.
\end{aligned}\ea
Under the leading-order correction of $c_i$, we can expand the solution to
\ba\begin{aligned}
&f_0(r)=1-\frac{m}{r}+\frac{q^2}{2 r^2}+f_1(r)+\math{O}(c_i^2)\,,\\
\m_0(r)=1+&\m_1(r)+\math{O}(c_i^2
)\,,\,\Y_0(r)=-\frac{q}{r}+\Y_1(r)+\math{O}(c_i^2)\,,
\end{aligned}\ea
where we used the fact that the background spacetime is a Reissner-Nordstrom black hole solution. $f_1(r)$, $\m_1(r)$ and $\Y_1(r)$ are the linear functions of $c_i$.

From the EOM $\grad_a E_F^{ab}=0$, we can obtain
\ba\begin{aligned}
\Y_1(r)=&\frac{2q}{5r^5}[c_5 q^2+c_6(6q^2-5mr)+(8c_7+4c_8)q^2]\\
&+q\int \frac{\m_1(r)}{r^2}dr\,.
\end{aligned}\ea
Substituting the above result to $H^{vv}=0$, it is easy to obtain
\ba\begin{aligned}
\m_1(r)=\frac{q^2}{r^4}(c_2+4c_3+10 c_4+3 c_5+3 c_6)\,,
\end{aligned}\ea
which gives
\ba\begin{aligned}\label{At}
\Y_1(r)=&-\frac{q^3}{5r^2}\left[c_2+4c_3+10 c_4+c_5\right.\\
&\left.-\left(9-10mrq^{-2}\right)c_6-16 c_7-8 c_8\right]\,.
\end{aligned}\ea
This result shows a different expression to $A_a$ given by Eq. (6) of \cite{Chenprl}. Finally, using $H^{\q\q}=0$, we can find that $f(r)$ shows the same expression of $g_{tt}$ in \cite{Chenprl}. Therefore, in the first-order gedanken experiments, the condition of not destroying an extremal solution is also given by Eq. (14) in \cite{Chenprl}, i.e.,
\ba\begin{aligned}\label{condition}
\d m-\sqrt{2}\d q\left(1+\frac{4c_0}{5q^2}\right)\geq 0\,.
\end{aligned}\ea
with
\ba\begin{aligned}
c_0\equiv c_2+4c_3+c_5+c_6+4c_7+2c_8\,.
\end{aligned}\ea
The condition that the test particle can drop into the horizon or the infalling matter satisfies the null energy condition is given by Eqs. (18) and (27) of \cite{Chenprl}, i.e.,
\ba\begin{aligned}\label{ineq}
\d m-\F_H^c\d q\geq 0\,,
\end{aligned}\ea
in which $\F_H^c\equiv-\left.\x^aA_a\right|_H$ with $\x^a=(\pd/\pd v)^a$ is the electric potential of the black hole. Using our corrected expression \eq{At} of $A_a$, for the extremal black hole, we have
\ba\begin{aligned}\label{Phiex}
\F_H^\text{ext}=\sqrt{2}\left(1+\frac{4c_0}{5q^2}\right)+\math{O}(c_i^2)\,,
\end{aligned}\ea
which is different from Eq. (11) of \cite{Chenprl}. Then, the inequality \eq{ineq} shows the same expression as inequality \eq{condition}, which implies that the extremal charged black hole cannot be destroyed. This result is just contrary to that shown by Ref. \cite{Chenprl} where there are destructions of the extremal black holes. This implies that after correcting the solution, their letter \cite{Chenprl} cannot show the ability of WCCC to constrain the gravitational theories.


\end{document}